\documentclass[aps,prd,preprint,groupedaddress,floatfix]{revtex4-1}
\usepackage{amsfonts}
\usepackage{graphicx}

\def\endprf{\hfill  {\vrule height6pt width6pt depth0pt}\medskip}

\begin{document}

%\preprint{}

\title{Exact solutions for classical Yang-Mills fields}


\author{Marco Frasca}
\email[]{marcofrasca@mclink.it}
%\thanks{}
\affiliation{Via Erasmo Gattamelata, 3 \\ 00176 Roma (Italy)}


\date{\today}

\begin{abstract}
We provide a set of exact solutions of the classical Yang-Mills equations. They have the property to satisfy a massive dispersion relation and hold in all gauges. These solutions can be used to describe the vacuum of the quantum Yang-Mills theory and so, they provide a general framework to build a quantum field theory. The components of the field become separated on a generic gauge but are all equal just in the Lorenz (Landau) gauge.
\end{abstract}

%\begin{keyword}



%\end{keyword}

\maketitle

%\section{}
%\label{}



Having a set of exact solutions to a classical field theory can be a sound starting point to build the corresponding quantum theory. This happens as the classical solutions can represent the behavior of the vacuum expectation value when quantum corrections are neglected. This is what happens in the Higgs mechanism but it is a general approach in the way to guess the right quantum theory. Indeed, a number of similar solutions are known \cite{Actor:1979in}. It is generally difficult to solve Yang-Mills equations being nonlinear and general methods to solve nonlinear equations beyond perturbation theory are not known.

Some years ago \cite{Frasca:2007uz} we proposed an approach to solve the Yang-Mills theory in the low-energy limit starting from a set of classical solution initially proposed by Smilga \cite{Smilga:2001ck}. Smilga solutions are rather peculiar as they correspond to the homogeneous case (no dependence on space variables) and are equal for all the components of the Yang-Mills field. They appear suitable for a gradient expansion of the classical field equations, what we did in \cite{Frasca:2007uz}. In that paper we put forward a theorem that mapped scalar field solutions to Yang-Mills solutions. This theorem was criticized by Terence Tao and, after we fixed it in \cite{Frasca:2009yp,tao2009}, it seemed that this mapping could be considered to hold only in an asymptotic sense when the coupling is taken to run to infinity, with the notable exception of the Lorenz (Landau) gauge where it is exact. This difficulty was harmless for the conclusions given in the original paper \cite{Frasca:2007uz} but we were left with the idea that the mapping could be exact instead, as the modifications introduced by the gauge fixing term do not appear to modify the equations of motion too much.

In this letter we show that things stay in this way obtaining the exact solutions in the most general case of any gauge. Anyhow, there is a substantial difference from a trivial expectation that keeps the Tao's argument sound. The condition of all equal components does not apply in the most general case but can be easily recovered in the limit of the coupling running to infinity as proved in \cite{Frasca:2009yp}.

Below, we will follow a step by step derivation making of this letter the proper companion to our preceding works and completing the argument on the behavior of the Yang-Mills theory and the scalar field theory in the low-energy limit. This represents a generalization of the solutions proposed by Smilga in \cite{Smilga:2001ck}.

To start, let us write down the equations of motion for the Yang-Mills field
\begin{equation}
\partial^\mu\partial_\mu A^a_\nu-\left(1-\frac{1}{\alpha}\right)\partial_\nu(\partial^\mu A^a_\mu)+gf^{abc}A^{b\mu}(\partial_\mu A^c_\nu-\partial_\nu A^c_\mu)+gf^{abc}\partial^\mu(A^b_\mu A^c_\nu)+g^2f^{abc}f^{cde}A^{b\mu}A^d_\mu A^e_\nu = 0
\end{equation}
that specialize to SU(2) for the sake of simplicity, with $f^{abc}=\varepsilon_{abc}$ the Levi-Civita symbol, as
\begin{equation}
\partial^\mu\partial_\mu A^a_\nu-\left(1-\frac{1}{\alpha}\right)\partial_\nu(\partial^\mu A^a_\mu)+g\varepsilon_{abc}A^{b\mu}(\partial_\mu A^c_\nu-\partial_\nu A^c_\mu)+g\varepsilon_{abc}\partial^\mu(A^b_\mu A^c_\nu)+g^2\varepsilon_{abc}\varepsilon_{cde}A^{b\mu}A^d_\mu A^e_\nu = 0.
\end{equation}
We use the following formula to simplify
%\begin{equation}
%\varepsilon_{ijk}\varepsilon_{lmn} = 
%\delta_{il}\left( \delta_{jm}\delta_{kn} - \delta_{jn}\delta_{km}\right) - \delta_{im}\left( \delta_{jl}\delta_{kn} - %\delta_{jn}\delta_{kl} \right) + \delta_{in} \left( \delta_{jl}\delta_{km} - \delta_{jm}\delta_{kl} \right).
%\end{equation}
%and
\begin{equation}
\varepsilon_{ijk}\varepsilon_{imn} = \delta_{jm}\delta_{kn} - \delta_{jn}\delta_{km}.
\end{equation}
This will yield
\begin{eqnarray}
&&\partial^\mu\partial_\mu A^a_\nu-\left(1-\frac{1}{\alpha}\right)\partial_\nu(\partial^\mu A^a_\mu)+g\varepsilon_{abc}A^{b\mu}(\partial_\mu A^c_\nu-\partial_\nu A^c_\mu)+g\varepsilon_{abc}\partial^\mu(A^b_\mu A^c_\nu) \nonumber \\
&&+g^2(A^{e\mu}A^a_\mu A^e_\nu-A^{b\mu}A^b_\mu A^a_\nu) = 0.
\end{eqnarray}
But it is $A_\mu^a=(A_\mu^1,A_\mu^2,A_\mu^3)$ and this gives
\begin{eqnarray}
&&\partial^\mu\partial_\mu A^1_\nu-\left(1-\frac{1}{\alpha}\right)\partial_\nu(\partial^\mu A^1_\mu)+ \nonumber \\
&&gA^{2\mu}(\partial_\mu A^3_\nu-\partial_\nu A^3_\mu)-gA^{3\mu}(\partial_\mu A^2_\nu-\partial_\nu A^2_\mu)+
g\partial^\mu(A^2_\mu A^3_\nu)-g\partial^\mu(A^3_\mu A^2_\nu)+ \nonumber \\
&&g^2(A^{2\mu}A^1_\mu A^2_\nu+A^{3\mu}A^1_\mu A^3_\nu-A^{2\mu}A^2_\mu A^1_\nu-A^{3\mu}A^3_\mu A^1_\nu) = 0 \nonumber \\
&&\partial^\mu\partial_\mu A^2_\nu-\left(1-\frac{1}{\alpha}\right)\partial_\nu(\partial^\mu A^2_\mu)+ \nonumber \\
&&gA^{3\mu}(\partial_\mu A^1_\nu-\partial_\nu A^1_\mu)-gA^{1\mu}(\partial_\mu A^3_\nu-\partial_\nu A^3_\mu)+
g\partial^\mu(A^3_\mu A^1_\nu)-g\partial^\mu(A^1_\mu A^3_\nu)+ \nonumber \\
&&g^2(A^{1\mu}A^2_\mu A^1_\nu+A^{3\mu}A^2_\mu A^3_\nu-A^{1\mu}A^1_\mu A^2_\nu-A^{3\mu}A^3_\mu A^2_\nu) = 0 \nonumber \\
&&\partial^\mu\partial_\mu A^3_\nu-\left(1-\frac{1}{\alpha}\right)\partial_\nu(\partial^\mu A^3_\mu)+ \nonumber \\
&&gA^{1\mu}(\partial_\mu A^2_\nu-\partial_\nu A^2_\mu)-gA^{2\mu}(\partial_\mu A^1_\nu-\partial_\nu A^1_\mu)+
g\partial^\mu(A^1_\mu A^2_\nu)-g\partial^\mu(A^2_\mu A^1_\nu)+ \nonumber \\
&&g^2(A^{1\mu}A^3_\mu A^1_\nu+A^{2\mu}A^3_\mu A^2_\nu-A^{1\mu}A^1_\mu A^3_\nu-A^{2\mu}A^2_\mu A^3_\nu) = 0.
\end{eqnarray}
Let us now put \cite{Smilga:2001ck} $A_1^1=A_2^2=A_3^3=\phi$. The set collapses on the equations
\begin{eqnarray}
&&\partial^2\phi-\left(1-\frac{1}{\alpha}\right)\partial_1(\partial^1\phi)+2g^2\phi^3 = 0 \nonumber \\
&&\partial^2\phi-\left(1-\frac{1}{\alpha}\right)\partial_2(\partial^2\phi)+2g^2\phi^3 = 0 \nonumber \\
&&\partial^2\phi-\left(1-\frac{1}{\alpha}\right)\partial_3(\partial^3\phi)+2g^2\phi^3 = 0
\end{eqnarray}
but this is possible only for $\alpha=1$. This shows how the choice of the Landau gauge simplifies computations. But we can also take $A_1^1\ne A_2^2\ne A_3^3\ne 0$ and all other being 0. This will yield the following set of equations
\begin{eqnarray}
&&\partial^2A^1_1-\left(1-\frac{1}{\alpha}\right)\partial_1(\partial^1 A^1_1)+g^2[(A^2_2)^2+(A^3_3)^2] A^1_1 = 0 \nonumber \\
&&\partial^2A^2_2-\left(1-\frac{1}{\alpha}\right)\partial_2(\partial^2 A^2_2)+g^2[(A^1_1)^2+(A^3_3)^2] A^2_2 = 0 \nonumber \\
&&\partial^2A^3_3-\left(1-\frac{1}{\alpha}\right)\partial_3(\partial^3 A^3_3)+g^2[(A^1_1)^2+(A^2_2)^2] A^3_3 = 0.
\end{eqnarray}
We have shown that in this case the solutions change from exact to asymptotic for $g\rightarrow\infty$ \cite{Frasca:2009yp}. We would like to follow a different approach and get the exact solution. So, given the {\sl ansatz}
\begin{eqnarray}
A^{1}_1&=&X\cdot{\rm sn}(p\cdot x,-1) \nonumber \\
A^{2}_2&=&Y\cdot{\rm sn}(p\cdot x,-1) \nonumber \\ 
A^{3}_3&=&Z\cdot{\rm sn}(p\cdot x,-1)
\end{eqnarray}
with the dispersion relation $p^2=\mu^2g$ to hold and $\mu$ an integration constant with the dimension of an energy, we get the following set of algebraic equations
\begin{eqnarray}
    Y^2+Z^2&=&\frac{2}{g^2}\left(1-\frac{1}{\alpha}\right)p_1^2+\mu^2\frac{2}{g} \nonumber \\
		X^2+Z^2&=&\frac{2}{g^2}\left(1-\frac{1}{\alpha}\right)p_2^2+\mu^2\frac{2}{g} \nonumber \\
		X^2+Y^2&=&\frac{2}{g^2}\left(1-\frac{1}{\alpha}\right)p_3^2+\mu^2\frac{2}{g}
\end{eqnarray}
that is easily solved. This shows that the idea of an asymptotic mapping in \cite{Frasca:2009yp} was correct as the contributions that select the different components goes like $O(1/g^2)$ and so negligible in the limit $g\rightarrow\infty$. But here we proved that such a mapping is indeed exact provided the proper solution for the given gauge is used. For $\alpha=1$ (Landau gauge) one has the expected result \cite{Frasca:2009yp}
\begin{equation}
   A_1^1=A_2^2=A_3^3=\frac{\mu}{(2g^2)^\frac{1}{4}}\cdot{\rm sn}(p\cdot x,-1).
\end{equation}
A different gauge removes the degeneracy of the components. This argument can be extended to any gauge group without difficulty other than a complication into algebraic computations. The relevance of these solutions relies on the particle-like dispersion relation that satisfy. This appears proper to a massive field even if we started from a massless theory. This is peculiar to a nonlinear field theory as we already proved for the scalar field \cite{Frasca:2009bc,Frasca:2013tma}. This means that, at a pure classical level, a massless $\phi^4$ theory can share identical solutions with the classical Yang-Mills field theory and these have a massive dispersion relation. In general we could write for this case
\begin{equation}
    A_\kappa^\kappa(x)=f_k(p,g)\phi(x)
\end{equation}
with $f_k$ a function originating from the gauge choice, the replacement $\lambda\rightarrow Ng^2$ for a generic SU(N) theory and $\phi=\mu(2/\lambda)^\frac{1}{4}\cdot{\rm sn}(p\cdot x,-1)$, provided $p^2=\mu^2\sqrt{\lambda/2}$. In this way, these solution can be used to describe the vacuum state of the quantum field theory. Such a theory would develop a mass gap due to the dispersion relation\cite{Frasca:2007uz,Frasca:2011bd,Frasca:2013tma}. In such a case one observes the breaking of conformal symmetry being the vacuum expectation value not trivial.

In conclusion, we have derived a set of exact classical solutions of the Yang-Mills theory. These solutions show the property to satisfy a mass-like dispersion relation even if we started with a massless theory. They can be used to describe the vacuum of the corresponding quantum field theory that, in this way, will acquire a mass gap. This will be the content of a future work.



\end{document}